\documentclass{ptephy}

\preprintnumber{XXXX-XXXX} 

\usepackage[normalem]{ulem}  
\usepackage{color} 

\renewcommand\sout{\bgroup \color[rgb]{1,0,0} \ULdepth=-.5ex \ULset}

\newcommand{\Slash}[1]{{\ooalign{\hfil/\hfil\crcr$#1$}}}
\newcommand{\tr}{{\rm tr}}
\newcommand{\eorep}{{\eta^{(\prime)}}}

\newcommand{\so}{{\left<\sigma_0\right>}}
\newcommand{\soo}{{\left<\sigma_8\right>}}

\newcommand{\pcm}{p_{cm}}

\usepackage{graphicx}	
\allowdisplaybreaks[1]

\begin{document}

\title{Investigation of the $\eta'N$ system using the linear sigma
model}

\author{\name{Shuntaro Sakai}{1,\ast}\footnote{Present address: Research
Center for Nuclear Physics (RCNP), Osaka University, Ibaraki, Osaka,
567-0047, Japan} and \name{Daisuke Jido}{2}}

\address{\affil{1}{Department of Physics, Kyoto University,
Kitashirakawa-Oiwakecho, Kyoto 606-8502, Japan}
\affil{2}{Department of Physics, Tokyo Metropolitan University, Hachioji
192-0397, Japan}
\email{shsakai@rcnp.osaka-u.ac.jp}}

\begin{abstract}%
 In this paper, we investigate the $\eta'N$ system using the
 three-flavor linear
 sigma model including
 the effect of
 flavor SU(3) symmetry
 breaking.
 The $\eta'N$ bound state is also found in the case including
 flavor symmetry breaking and coupling with the $\eta N$
 and $\pi N$ channels.
 The $\eta'N$ interaction becomes more attractive with the
 inclusion of flavor symmetry breaking,
 which causes mixing between the singlet and octet scalar mesons.
 The existence of such a bound state would have some impact on the
 $\eta'$-nucleus system, which is of interest from both
 theoretical and experimental viewpoints.
\end{abstract}

\subjectindex{D32}

\maketitle

\section{Introduction\label{sec_intro}}
The properties of the $\eta'$ meson have
attracted continuous attention
in hadron physics.
Its relatively large mass
compared with other low-lying pseudoscalar
mesons, such as $\pi$ or $\eta$, is
summarized as the U$_A$(1) problem \cite{Weinberg:1975ui}, and many
studies~\cite{'tHooft:1986nc,Christos:1984tu,Witten:1979vv,Veneziano:1979ec}
have been devoted to
the role of the U$_A$(1)
anomaly in quantum chromodynamics (QCD)~\cite{Adler:1969gk,Bell:1969ts,Bardeen:1969md}.
In the three-flavor system,
{on the other hand,} the essential role of
chiral symmetry breaking in cooperation with the U$_A$(1) anomaly for the
generation of the $\eta'$ mass is pointed out in
Refs.~\cite{Pisarski:1983ms,Kikuchi:1987jr,Kunihiro:1987bb,Lee:1996zy,Cohen:1996ng,Jido:2011pq}.

Partial restoration of chiral symmetry in the nuclear medium
attracts our interest;
the magnitude of the order parameters of chiral symmetry breaking is
expected to be reduced in the nuclear
medium~\cite{Drukarev:1991fs,Cohen:1991nk,Brockmann:1996iv},
and its possible effects on hadronic phenomena have been studied
intensively (see Ref.~\cite{Hayano:2008vn} for a recent review).
In particular, analyses of the pion$-$nucleus system suggest that such
a reduction of the order parameter in the nuclear medium actually occurs
with about 30\% suppression of the size of the quark condensate at the
normal nuclear
density~\cite{Suzuki:2002ae,Friedman:2004jh,Kolomeitsev:2002gc,Jido:2008bk}.

Taking account of the partial restoration of chiral symmetry and the
strong connection of the $\eta'$ mass and chiral symmetry breaking, we
expect that the mass of the $\eta'$ meson is reduced in the nuclear
medium~\cite{Jido:2011pq}, though the in-medium mass of $\eta'$ is
concerned from the viewpoint of the effective restoration of the U$_A$(1)
symmetry~\cite{Pisarski:1983ms,Kikuchi:1987jr,Kunihiro:1989my,Kapusta:1995ww};
some model calculations suggest that the $\eta'$ mass reduces as much as
about 100 MeV at the normal nuclear density in association with
chiral restoration~\cite{Costa:2002gk,Nagahiro:2006dr,Sakai:2013nba}.
Then, we expect to find some information on chiral restoration in
the nuclear medium through investigation of the $\eta'$-nucleus
system studied from both theoretical and experimental
aspects~\cite{Tsushima1,Tsushima2,Tsushima3,Nagahiro:2004qz,Nagahiro:2011fi,Nanova:2012vw,Itahashi:2012ut,Nagahiro:2012aq,Nanova:2013fxl}.

For the study of the in-medium properties of the $\eta'$ meson,
the interaction between $\eta'$ and the nucleon
$N$ is a basic piece.
The $\eta'N$ system is theoretically studied in
Refs.~\cite{Bass:2005hn,Kawarabayashi:1980uh,Borasoy:1999nd,Oset:2010ub}.
Experimentally, analysis of $\pi^-p\rightarrow\eta'n$ near the
$\eta'n$ threshold suggests the existence of the narrow and shallow
$\eta'n$ bound state \cite{Moyssides:1983}, and a threshold enhancement
is seen in the $\eta'$ photoproduction \cite{Kashevarov:2016owq}.
On the other hand, recent analysis of the $pp\rightarrow pp\eta'$
reaction suggests a small scattering
length~\cite{Moskal:2000gj,Moskal:2000pu,Czerwinski:2014yot}.

In our previous study, we investigated the mass of the $\eta'$ meson in
the nuclear medium and the $\eta'N$ two-body interaction in free space
using the three-flavor linear sigma model with
nucleon~\cite{Sakai:2013nba,Sakai:2014zoa}.
Within the leading order of the momentum expansion,
we found that the attractive $\eta'N$ interaction is
induced by the scalar meson exchange;
the $\sigma\eta'\eta'$ coupling is induced by U$_A$(1) symmetry
breaking, and the transition to the $\eta N$ channel is suppressed by
the large mass of the octet scalar meson.
The $\eta'N$ interaction is strong enough to form the $\eta'N$ bound
state.
The possible appearance of the bound-state signal in the
$\eta'$ photoproduction process
off the deuteron is investigated in Ref.~\cite{Sekihara:2016jgx}.

In this study, we investigate the $\eta'N$ system based on the linear
sigma model.
Due to the lack of phenomenological information of the $\eta'N$
interaction, we make good use of theoretical considerations and symmetry
properties of hadrons.
The linear sigma model is one of the well-contained effective models by
chiral symmetry and has the mechanism of chiral restoration.
In the exploratory stage, it would be interesting to see what comes out from
such theoretical analyses.
Certainly, once we have experimental information on the $\eta'N$
interaction, we should go to the nonlinear realization of chiral
symmetry, in which more phenomenological analyses can be done based on
chiral effective theories.

In this paper, we evaluate the $\eta'N$ interaction including the
$\eta N$ and $\pi N$ transition at the $\eta'N$ threshold energy,
and investigate a possible nucleon resonance dynamically generated from
these $\eta'N$, $\eta N$, and $\pi N$ channels around the $\eta'N$
threshold.
A new finding of this study is that transition between the $\eta'N$ and
$\eta N$ or $\pi N$ channels are induced by the effect of the U(1)$_A$
anomaly in addition to the $\eta'N$ elastic channel in this setup.
We also take account of the effects of scalar meson mixing caused by the
explicit breaking of the flavor SU(3) symmetry, which was not taken into
account in the previous work.
The mixing induces the octet sigma meson exchange in the $\eta'N$
interaction.
We also calculate the mass and width of the possible state of  $\eta'N$
by including the scattering channels of $\eta N$ and $\pi N$.

This paper is constructed as follows:
the model used in this calculation is introduced in
Sect.~\ref{sec_model_setup}.
The result of the study is given in Sect.~\ref{sec_result}.
Finally, we give a summary and discussion in Sect.~\ref{sec_summary}.

\section{Model setup\label{sec_model_setup}}
In this section, we introduce the model used in this study,
that is
the three-flavor linear sigma model.
The Lagrangian is given as follows:
\begin{align}
 \mathcal{L}=&\frac{1}{2}\tr (\partial_\mu M\partial^\mu
 M^\dagger)-\frac{\mu^2}{2}\tr (MM^\dagger)
 -\frac{\lambda}{4}\tr(MM^\dagger)^2-\frac{\lambda'}{4}\left[\tr(MM^\dagger)\right]^2+A\tr(\chi
 M^\dagger+\chi^\dagger M)\notag\\
 &+\sqrt{3}B(\det M+\det M^\dagger) \notag\\
 &+\bar{N}\left[i\Slash{\partial}
 -m_N
 -g\left\{\left(\frac{
 \tilde{\sigma}_0}{\sqrt{3}}{\bf
 1}+\frac{\vec{\sigma}\cdot\vec{\tau}}{\sqrt{2}}+\frac{
 \tilde{\sigma}_8}{\sqrt{6}}{\bf
 1}\right)+i\gamma_5\left(\frac{\eta_0}{\sqrt{3}}{\bf 1}+\frac{\vec{\pi}\cdot\vec{\tau}}{\sqrt{2}}+\frac{\eta_8}{\sqrt{6}}{\bf 1}\right)\right\}\right]N,\label{eq_lag}
\end{align} 
where
\begin{align}
 M=&M_s+M_{ps},\\
 M_s=&\begin{pmatrix}
 \frac{\sigma_0}{\sqrt{3}}+\frac{\sigma_3}{\sqrt{2}}+\frac{\sigma_8}{\sqrt{6}}&a^+&\kappa^+\\
	  a^-&\frac{\sigma_0}{\sqrt{3}}-\frac{\sigma_3}{\sqrt{2}}+\frac{\sigma_8}{\sqrt{6}}
 & \kappa^0\\
 \kappa^-&\bar{\kappa}^0 &\frac{\sigma_0}{\sqrt{3}}-\sqrt{\frac{2}{3}}\sigma_8
 \end{pmatrix},\\
 M_{ps}=&\begin{pmatrix}
 \frac{\eta_0}{\sqrt{3}}+\frac{\pi_3}{\sqrt{2}}+\frac{\eta_8}{\sqrt{6}}&\pi^+&K^+\\
	  \pi^-&\frac{\eta_0}{\sqrt{3}}-\frac{\pi_3}{\sqrt{2}}+\frac{\eta_8}{\sqrt{6}}
 & K^0\\
 K^-&\bar{K}^0 &\frac{\eta_0}{\sqrt{3}}-\sqrt{\frac{2}{3}}\eta_8
	 \end{pmatrix},\\
 N=&^t(p,n),\\
 \chi=&\sqrt{3}{\rm diag}(m_u,m_d,m_s)=\sqrt{3}{\rm diag}(m_q,m_q,m_s).
\end{align}
The Gell-Mann matrix $\lambda_a$ and Pauli matrix
$\tau_a$
satisfy
the normalization $\tr(\lambda_a\lambda_b)=2\delta_{ab}$ and
$\tr(\tau_a\tau_b)=2\delta_{ab}$, respectively.
$\tilde{\sigma}_i$ $(i=0,8)$ means the fluctuation from its
mean field which will be explained later.
The meson and baryon fields belong to the $({\bf 3},\bar{\bf
3})\oplus(\bar{\bf 3},{\bf 3})$ representation of
SU(3)$_L\times$SU(3)$_R$ and
the Lagrangian is constructed to be invariant under the chiral
transformation.
The isospin symmetry is implemented by the degenerate $u$ and $d$ quark
masses $m_q=m_u=m_d$, and the flavor symmetry breaking appears from the
non-degenerate strange quark mass, $m_s\neq m_q$.
For the baryonic part of the Lagrangian, the irrelevant hyperons in this
study are omitted.

The vacuum is determined so as to minimize the effective potential
obtained with the tree-level approximation in this study.
The effective potential $V_\sigma(\sigma_0,\sigma_8)$ is given as
\begin{align}
 &V_\sigma(\sigma_0,\sigma_8)=\frac{\mu^2}{2}(\sigma_0^2+\sigma_8^2)+\frac{\lambda}{12}(\sigma_0^4+6\sigma_0^2\sigma_8^2-2\sqrt{2}\sigma_0\sigma_8^3+\frac{3}{2}\sigma_8^4)+\frac{\lambda'}{4}(\sigma_0^2+\sigma_8^2)^2\notag\\
 &\hspace{1cm}-\frac{2}{3}B(\sigma_0+\frac{\sigma_8}{\sqrt{2}})^2(\sigma_0-\sqrt{2}\sigma_8)-2A(2m_q(\sigma_0+\frac{\sigma_8}{\sqrt{2}})+m_s(\sigma_0-\sqrt{2}\sigma_8)).
\end{align}
The order parameter of the chiral symmetry breaking is the expectation
value of the neutral scalar meson fields $\sigma_i$ $(i=0,8)$.
The expectation value of $\sigma_3$ vanishes due to the isospin
symmetry.
Here, we decompose them into the mean field value
$\left<\sigma_i\right>$ and fluctuation $\tilde{\sigma}_i$;
$\sigma_i=\left<\sigma_i\right>+\tilde{\sigma}_i$.
The minimum conditions of the effective potential $\partial
V_\sigma/\partial\sigma_{0,8}=0$ are
satisfied for the expectation values of $\sigma_{0,8}$.
The expectation values are related to the meson decay constants;
at the tree level
they are given as
$f_\pi=\sqrt{2/3}\so+\soo/\sqrt{3}$ and
$f_K=\sqrt{2/3}\so-\soo/2\sqrt{3}$.
The non-zero vacuum expectation value $\so$ is responsible for the
spontaneous breaking of chiral symmetry in the linear sigma model, while
the finite vacuum expectation value $\soo$ is caused by the explicit
flavor symmetry breaking due to the non-degenerate strange quark mass.

The parameters contained in the meson part and
$\so$ and $\soo$ are determined to reproduce the observed
masses and decay constants of mesons.
The parameter in the nucleon part $g$ is fixed in order for the
magnitude of the quark
condensate to reduce by 35\% at the normal nuclear density, which is
suggested from analysis of the deeply bound state of the pionic
atom~\cite{Suzuki:2002ae}.
Details of the determination of the parameters are given in
Ref.~\cite{Sakai:2013nba}.
The input and the determined parameters are
given in Tables~\ref{tab_input_prmt} and \ref{tab_determined_prmt},
respectively.
The masses of the mesons used as input parameters come from the
values in Ref.~\cite{PDG}.
The masses of the $\pi$ and $K$ mesons are obtained with the isospin
average.
\begin{table}
 \begin{center}
  \caption{The input parameters for the determination of the parameters
  in the Lagrangian.
  These values are taken from Ref.~\cite{PDG}.
  The masses of $\pi$ and $K$ are isospin-averaged values.}
  \begin{tabular}{cccc}
   $f_\pi$ [MeV]& $f_K$ [MeV]& $m_\pi$ [MeV] &$m_K$ [MeV]\\
   $92.2$ & $110.4$ &$138.04$&$495.64$\\\hline
   &$m_\eta^2+m_{\eta'}^2$ [MeV$^2$]&$m_\sigma$ [MeV]\\
   &$547.85^2+957.78^2$ &$700$
  \end{tabular}
  \label{tab_input_prmt}
 \end{center}
\end{table}
\begin{table}
 \begin{center}
  \caption{The determined parameters in the Lagrangian.}
   \begin{tabular}{cccc}
    $\so$ [MeV]&$\soo$ [MeV]&$\mu^2$ [MeV$^2$]&$\lambda$ [-]\\
    $127.78$&$-21.02$ &$4.21\times 10^{4}$ &$58.80$ \\\hline
    $\lambda'$ [-]&$A$ [MeV$^2$]&$B$ [MeV]&$g$ [-]\\      
    $2.46$ &$7.17\times 10^4$ &$997.95$ &$9.84$ \\ 
   \end{tabular}
  \label{tab_determined_prmt}
 \end{center}
\end{table}
 
Here, we treat the mass of the $\sigma$ meson $m_\sigma$ as an input parameter.
The dependence on $m_\sigma$ of our result will be discussed later.

The mass eigenstates of the neutral
mesons are different from the eigenstates of
the flavor appearing in the Lagrangian (\ref{eq_lag})
due to the flavor symmetry breaking
from the difference between $m_q$ and $m_s$ in the Lagrangian.
The mass eigenstates of the isospin singlet scalar
(pseudoscalar) mesons are denoted by
$\sigma$ and $f_0$ ($\eta'$ and $\eta$).
They are related by
\begin{align}
 \begin{pmatrix}
  \sigma\\
  f_0
 \end{pmatrix}=&
 \begin{pmatrix}
  \cos\theta_s&\sin\theta_s\\
  -\sin\theta_s&\cos\theta_s 
 \end{pmatrix}
 \begin{pmatrix}
  \sigma_0\\
  \sigma_8
 \end{pmatrix},\label{eq_sigma_mixing_mtrx}\\
 \begin{pmatrix}
  \eta'\\
  \eta
 \end{pmatrix}
 =&
 \begin{pmatrix}
  \cos\theta_{ps}&\sin\theta_{ps}\\
  -\sin\theta_{ps}&\cos\theta_{ps}
 \end{pmatrix}
 \begin{pmatrix}
  \eta_0\\
  \eta_8
 \end{pmatrix},\label{eq_eta_mixing_mtrx}
\end{align}
where $\theta_{s}$ and $\theta_{ps}$ are the mixing angles in the scalar
and pseudoscalar sectors, respectively.
The matrices (\ref{eq_sigma_mixing_mtrx}) and
(\ref{eq_eta_mixing_mtrx}) diagonalize the mass matrices
written in
the bases of $\sigma_{0,8}$ and $\eta_{0,8}$, respectively.
The mass matrices are related by
\begin{align}
 (\sigma,f_0)
 \begin{pmatrix}
  m_{\sigma}^2&0\\
  0&m_{f_0}^2 
 \end{pmatrix}
 \begin{pmatrix}
  \sigma\\
  f_0
 \end{pmatrix}=&
 (\sigma_0,\sigma_8)
 \begin{pmatrix}
  m_{\sigma_0}^2&m_{\sigma_0\sigma_8}^2\\
  m_{\sigma_0\sigma_8}^2&m_{\sigma_8}^2 
 \end{pmatrix}
 \begin{pmatrix}
  \sigma_0\\
  \sigma_8
 \end{pmatrix},\\
 (\eta',\eta)
 \begin{pmatrix}
  m_{\eta'}^2&0\\
  0&m_{\eta}^2 
 \end{pmatrix}
 \begin{pmatrix}
  \eta'\\
  \eta
 \end{pmatrix}=&
 (\eta_0,\eta_8)
 \begin{pmatrix}
  m_{\eta_0}^2&m_{\eta_0\eta_8}^2\\
  m_{\eta_0\eta_8}^2&m_{\eta_8}^2 
 \end{pmatrix}
 \begin{pmatrix}
  \eta_0\\
  \eta_8
 \end{pmatrix}.
\end{align}
The explicit form of the meson mass is given in
Appendix~\ref{app_meson_mass}.
The off-diagonal part of the matrices, which causes the transition
between the singlet and octet mesons, appears from the explicit flavor
symmetry breaking.
The mixing angles are found as $\theta_s=21.0^\circ$ and
$\theta_{ps}=-6.2^\circ$ in the case of $m_\sigma=700$ MeV.
The masses of the scalar mesons, $\sigma$, $f_0$, and $a_0$ are
obtained as $700$, $1286$, and $1071$ MeV, respectively.

\section{Results\label{sec_result}}
In this section, we show the results of this study.
We first show the $\eta'N$
interaction and then
the $T$ matrix of $\eta'N$ with the coupling to the
$\eta N$ and $\pi N$ channels including the effect of the flavor
SU(3) symmetry breaking.
The effect of the flavor symmetry breaking appears from the nonzero
value of $\soo$.
This leads to the modification of the meson coupling and the mixing
property of mesons.
While the $\pi\pi N$ can give significant effects in general, we
omit it following the discussion given in Ref.~\cite{Oset:2010ub};
the calculation in Ref.~\cite{Inoue:2001ip} suggests the minor
correction for the $I=1/2$ channel amplitude due to the $\pi\pi N$
channel, while the $\eta'N$ channel is not included in the study.

\subsection{$\eta'N$ interaction\label{sec_epn_int}}
\begin{figure}[t]
 \centering
 \includegraphics[width=8.5cm]{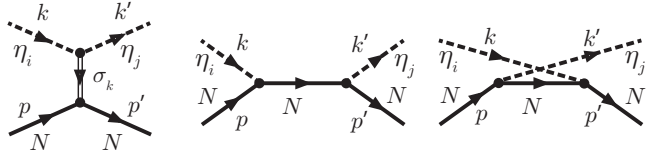}
 \caption{Diagrams taken into account in this study.
 The solid, double solid, and dashed lines mean the nucleon, scalar
 meson, and pseudoscalar meson.
 $\eta_{i,j}=\pi$, $\eta$, $\eta'$ and $\sigma_k=\sigma$, $f_0$, $a_0$.}
 \label{fig_epn_diag_tree}
\end{figure}

First, we evaluate the $\eta'N$,
$\eta N$, and $\pi N$
interactions.
Here, the matrix element of the interaction is evaluated within the tree-level approximation.
The diagrams included for the evaluation of the $\eta'N$ interaction are
shown in Fig.~\ref{fig_epn_diag_tree}.
Here, $\eta_i$ and $\eta_j$ are the mesons in the initial and final states
$(\eta_{i,j}=\pi,\eta,\eta')$.
The matrix elements are
calculated as follows:
\begin{align}
 &\mathcal{M}_{\eta_iN\rightarrow\eta_jN}
 =\bar{u}(p',s')\left[\frac{g_{\sigma_k\eta_i\eta_j}g_{\sigma_kN}}{q^2-m_{\sigma_k}^2+i\epsilon}+\frac{g_{\eta_iN}g_{\eta_jN}\Slash{k}}{(p+k)^2-m_N^2+i\epsilon}-\frac{g_{\eta_iN}g_{\eta_jN}\Slash{k}'}{(p-k')^2-m_N^2+i\epsilon}\right]u(p,s),\label{eq_mat_elem_ij}
\end{align}
where $p$ $(p')$, and $k$ $(k')$ are the momenta of the incoming
(outgoing) nucleon and meson in the center-of-mass frame, respectively,
which are explicitly written as
\begin{align}
 &p^\mu=(E_N=\sqrt{\pcm^2+m_N^2},0,0,\pcm),\label{eq_pmu}\\
 &k^\mu=(E_i=\sqrt{\pcm+m_i^2},0,0,-\pcm),\\
 &p'^\mu=(E_N'=\sqrt{\pcm'^2+m_N^2},\pcm'\sin\theta,0,\pcm'\cos\theta),\\
 &k'^\mu=(E_j=\sqrt{\pcm'^2+m_j^2},-\pcm'\sin\theta,0,-\pcm'\cos\theta),\label{eq_kpmu}
\end{align}
where
{$m_i$ and $m_j$ are the masses of the pseudoscalar mesons
$\eta_i$ and $\eta_j$, respectively, and} $\pcm$ and $\pcm'$ are the momenta of hadrons
in the initial and final states
in the center-of-mass frame.
They
are
given by
$\pcm=\frac{1}{2W}\sqrt{\lambda(W^2,m_N^2,m_i^2)}$ and
$\pcm'=\frac{1}{2W}\sqrt{\lambda(W^2,m_N^2,m_j^2)}$ with
$\lambda(x,y,z)=x^2+y^2+z^2-2xy-2yz-2zx$ and the center-of-mass energy
$W=E_N+E_i$.
Here, we assume that the initial particles come along with the $z$-axis.
In Eq.~(\ref{eq_mat_elem_ij}),
$q$ denotes the
momentum transfer defined as
$q=p'-p$.
Here, we write the total momentum of the system as
$P^\mu=p^\mu+k^\mu=(W,0,0,0)$.
The coupling strengths of the hadrons, $\eta_i(\sigma_i)\bar{N}N$ and
$\sigma_k\eta_i\eta_j$,
are denoted as $g_{\eta_i(\sigma_i)N}$ and $g_{\sigma_k\eta_i\eta_j}$.
Their explicit expressions are given in Appendix~\ref{app_coupling_mesons}.

We solve the scattering equation with the $s$-wave interaction
deduced from the matrix element Eq.~(\ref{eq_mat_elem_ij}).
For this purpose, we introduce the $s$-wave interaction vertex
$V_{\eta_iN\rightarrow\eta_jN}$ defined by
\begin{align}
 V_{\eta_iN\rightarrow \eta_jN}=\frac{1}{2}\sum_{\rm
 spin}\frac{1}{2}\int_{-1}^1d(\cos\theta)\mathcal{M}_{\eta_iN\rightarrow
 \eta_jN}.\label{eq_proj_def}
\end{align}
Here, we have performed $s$-wave projection and taken spin average for
the initial nucleon and spin sum over the final nucleon.
The $s$-wave component would be dominant near the $\eta'N$ threshold
that we are focusing on.
Moreover,
we take
the $I=1/2$ component of the $\pi N$ channel, which can couple to the $\eta'N$ channel.
The explicit form of $V_{\eta_iN\rightarrow\eta_jN}$ is given in Appendix~\ref{app_vij}, and
the value at the $\eta'N$ threshold is presented in Table.~\ref{tab_val_vij}.
$V_{\eta_iN\rightarrow\eta_jN}$ is evaluated using the physical value of
the nucleon and pseudoscalar mesons.
\begin{table}
 \begin{center}
  \caption{Values of $V_{i\rightarrow j}$ at the $\eta'N$
  threshold.
  For the $\pi N$ channel, the projection into $I=1/2$ is done.}
  \begin{tabular}{ccc}
   $V_{\eta'N\rightarrow\eta'N}$ [MeV$^{-1}$]&$V_{\eta N\rightarrow\eta
       N}$ [MeV$^{-1}$]&$V_{\pi N\rightarrow\pi N}$ [MeV$^{-1}$]\\
   $-7.14\times 10^{-2}$&2.02$\times 10^{-2}$ &$-8.54\times
	   10^{-3}$\\\hline
   $V_{\eta'N\rightarrow\eta N}$ [MeV$^{-1}$]&$V_{\eta'N\rightarrow\pi
       N}$ [MeV$^{-1}$]&$V_{\eta N\rightarrow\pi N}$ [MeV]$^{-1}$\\
   9.92$\times 10^{-3}$ &$5.67\times 10^{-2}$
       &$1.74\times 10^{-2}$ \\
  \end{tabular}
  \label{tab_val_vij}
 \end{center} 
\end{table}

Comparing the value of the $s$-wave $\eta'N$ interaction obtained
in this calculation with that obtained in
Refs.~\cite{Sakai:2013nba,Sakai:2014zoa} where the flavor symmetry
breaking is not incorporated in the calculation of the $\eta'N$
interaction, we find that the present value of
$V_{\eta'N\rightarrow \eta'N}$ is larger than the previous
calculation, in which we have obtained
$V_{\eta'N\rightarrow\eta'N}=-0.054$ and
$V_{\eta'N\rightarrow\eta N}=0.012$ MeV$^{-1}$ which are evaluated also
without the meson mixing and under the momentum expansion.
As demonstrated in Refs.~\cite{Sakai:2013nba,Sakai:2014zoa}, the
scalar meson exchange has the leading contribution in the momentum
expansion in these channels.
This result implies that the SU(3) flavor symmetry breaking and the
resultant scalar meson mixing give substantial effects on the $\eta'N$
interaction.
In particular, the parameter $g$ which characterizes the coupling of the
$\sigma$ meson and nucleon is larger when we take account of the mixing
effect.

Next, we comment on the effect of the $\eta$-$\eta'$ mixing.
As we mentioned in the previous section, the mixing angle between $\eta'$
and $\eta$ is as small as $-6.2^\circ$.
If we omit the mixing between the $\eta$ and $\eta'$
mesons,
$V_{\eta'N\rightarrow\eta'N}$, $V_{\eta'N\rightarrow\eta N}$,
$V_{\eta'N\rightarrow\pi N}$, $V_{\eta N\rightarrow\eta N}$, and
$V_{\eta N\rightarrow\pi N}$ are
given as $-0.068$, $0.020$, $0.058$, $0.017$, and $0.011$ MeV$^{-1}$,
respectively,
where the mixing effect
does not affect
$V_{\pi N\rightarrow\pi N}$.
With the omission of the pseudoscalar mixing, the $\eta'$ and
$\eta$ correspond to the flavor eigenstates $\eta_0$ and $\eta_8$,
respectively.
The modification by the $\eta$-$\eta'$ mixing
appears to be small compared with the values in Table~\ref{tab_val_vij}.

\subsection{$\eta'N$ system\label{sec_epn_tmtrx}}
We evaluate the $T$ matrix of the $\eta'N$ system by solving the
scattering equation,
\begin{align}
 T_{i\rightarrow j}=V_{i\rightarrow j}+V_{i\rightarrow
 k}G_kT_{k\rightarrow j},
\end{align}
where $i$, $j$, and $k$ mean $\pi N$, $\eta N$, or $\eta'N$,
and the $s$-wave
projection and the spin average is used to obtain
the interaction kernel of the scattering equation $V_{i\rightarrow j}$.
Here, we use the value of $V_{i\rightarrow j}$ evaluated at the $\eta'N$
threshold to evaluate $T_{i\rightarrow j}$.
The scattering equation can be solved in an algebraic way:
$T_{i\rightarrow j}=[(1-VG)^{-1}V]_{ij}$.
The loop function $G_i$ in the equation which has
an ultraviolet divergence is regularized by the use of the dimensional regularization.
Then, $G_i$ is written as follows:
\begin{align}
 G_i(W)=&i\int\frac{d^4q}{(2\pi)^4}\frac{2m_N}{(P-q)^2-m_N^2+i\epsilon}\frac{1}{q^2-m_P^2+i\epsilon}\notag\\
 =&\frac{2m_N}{(4\pi)^2}\left[a_i(\mu)+\ln\left(\frac{m_N^2}{\mu^2}\right)+\frac{W^2-m_N^2+m_P^2}{2W^2}\ln\left(\frac{m_P^2}{m_N^2}\right)\right.\notag\\
 &+\frac{\pcm}{W}\left\{\ln(W^2-m_N^2+m_P^2+2\pcm
 W)+\ln(W^2+m_N^2-m_P^2+2\pcm W)\right.\notag\\
 &\left.\left.-\ln(-W^2-m_N^2+m_P^2+2\pcm W)-\ln(-W^2+m_N^2-m_P^2+2\pcm W)\right\}\right],
\end{align}
where $m_N$ and $m_P$ are the masses
of the nucleon and the pseudoscalar
meson, and $\mu$ is a renormalization point which is fixed as $\mu=m_N$
in this calculation.
The subtraction constant $a_i(\mu)$ is determined with the natural
renormalization scheme \cite{Hyodo:2008xr};
this implies that
the contribution from other degrees of freedom than those
contained in this model is excluded.
The values of the subtraction constants are $a_{\eta'N}(m_N)=-1.838$,
$a_{\eta N}(m_N)=-1.239$, and $a_{\pi N}(m_N)=-0.398$.

It is known that the linear sigma model is even less suitable for the
$\pi N$
system with energies far from the $\pi N$ threshold.
The present model may not control the $\pi N$ system well.
Actually, evaluating the cross section of $\pi^+n \rightarrow \eta' p$
using the $T$ matrix obtained in this model, we find that the model
substantially overestimates the transition cross section reported in
Ref.~\cite{Rader:1973mx};
the cross section is evaluated as about 1.3 mb, which is about ten
times larger than the value in Ref.~\cite{Rader:1973mx}.
To control the transition strength to the $\pi N$ channel, we scale down
the vertex $V_{\pi N\rightarrow\eta'N}$ so as to reproduce the
experimental data,
$\sigma_{\pi^+n\rightarrow\eta'p} = 0.1$ mb at $W=2$ GeV.
This can be achieved by multiplying a factor $x_{13} = 0.153$ to the vertex
$V_{\pi N\rightarrow\eta'N}$ for $m_\sigma = 700$ MeV.

To obtain the mass and width of the $\eta'N$ bound (or resonance) state,
we perform analytic continuation of the obtained $T$ matrix to the
complex energy plane and find a pole of the $T$ matrix. 
From the pole residue, we can obtain the coupling of the hadrons in the
channel $i$ and the bound state $g_{iN^\ast}$.
Near the pole of $T_{i\rightarrow j}$, we can write $T_{i\rightarrow j}$
as
\begin{align}
 T_{i\rightarrow j}=\frac{g_{iN^\ast}g_{jN^\ast}}{W-m_R}+({\rm regular\ part\ at\
 }W=m_R),
\end{align}
where the complex value $m_R$ denotes the pole position of the $T$ matrix.

\begin{figure}[t]
 \centering
 \includegraphics[width=8cm]{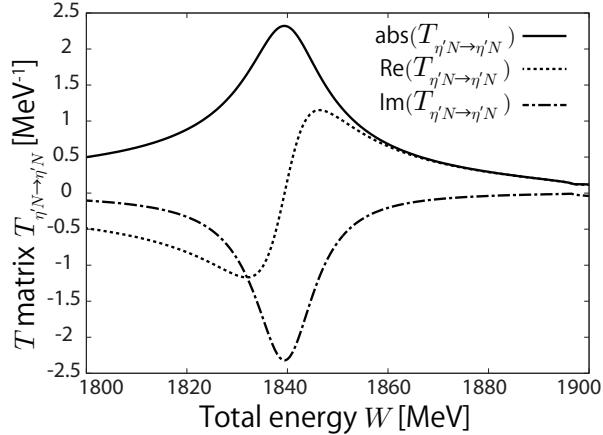}
 \caption{Plot of the absolute value, real part, and imaginary part of the
 $T$ matrix in the $\eta'N$ channel with $m_\sigma=700$ MeV.
 The solid, dashed, and dash-dotted lines
 stand for the absolute
 value, real part, and imaginary part of the
 $T$ matrix
 $T_{\eta'N\rightarrow \eta'N}$,
 respectively.}
 \label{fig_plot_tmtrx}
 \end{figure}

The binding energy, scattering length of $\eta'N$, and the
coupling of the bound state with the channel $i$,
$g_{iN^\ast}$, with $m_\sigma=700$ MeV are given in
Table~\ref{tab_val_bound_state}.
The definitions of the scattering length $a$ and effective range $r_e$
are the same as those in Ref.~\cite{Ikeda:2011dx}:
\begin{align}
 a=&\left.-\frac{m_N}{4\pi W}T(W)\right|_{W=m_N+m_P},\\
 r_e=&\left.\frac{d^2}{dk^2}\left(-\frac{m_N}{4\pi W}T(W)\right)^{-1}\right|_{W=m_N+m_P},
\end{align}
where the momentum of a meson in the center-of-mass frame $k$ and the
center-of-mass energy
$W$ are related by $k=\frac{1}{2W}\sqrt{[W^2-(m_N+m_P)^2][W^2-(m_N-m_P)^2]}$.
$m_N$ and $m_P$ are the masses of the nucleon and pseudoscalar meson, respectively.
\begin{table}
 \begin{center}
  \caption{The values of the pole position, scattering length, and
  couplings of the bound state with the channel $i$, $g_{iN^\ast}$, with
  $m_\sigma=700$ MeV.}
  \begin{tabular}{cc}
   Pole position [MeV]& Binding energy [MeV]\\
   $1839.7-7.2i$ & $57.0-7.2i$\\\hline
   Scattering length [fm]& Effective range [fm]\\
   $-0.98+5.7\times 10^{-2}i$&$0.22-6.8\times 10^{-3}i$ 
  \end{tabular}
  \begin{tabular}{ccc}\hline
   $g_{\eta'NN^\ast}$ [$-$]&$g_{\eta NN^\ast}$ [$-$]& $g_{\pi NN^\ast}$
	   [$-$]\\
   $4.1+0.15i$ &$-0.38+0.25i$& $-0.32+1.7\times 10^{-2}i$\\
  \end{tabular}
  \label{tab_val_bound_state}
 \end{center}
\end{table}
A plot of the absolute value, real part, and imaginary part of the $T$
matrix in the $\eta'N$ elastic channel $T_{\eta'N\rightarrow \eta'N}$ is
shown in Fig.~\ref{fig_plot_tmtrx}.
One can see a clear peak in the $\eta'N$ bound state in
$\left|T_{\eta'N\rightarrow\eta'N}\right|$.

In this model we can treat the mass of the $\sigma$ meson as an input
parameter, as mentioned in Sect.~\ref{sec_model_setup}.
In the chiral effective theory in the nonlinear realization,
the sigma meson mass should be irrelevant for low-energy dynamics.
Varying the mass of the sigma meson $m_\sigma$, we can find a pole
of the $\eta'N$ bound state.
The pole position and the scale factor $x_{13}$ for the fit of
$\sigma_{\pi^+n\rightarrow\eta'p}$ with the change of the $\sigma$
mass are given in Table~\ref{tab_pole_traj}.
\begin{table}[t]
 \begin{center}
  \caption{The pole position of the $\eta'N$ bound state
  varying the mass of the scalar meson $m_\sigma$.}
  \begin{tabular}{cccc}
   $m_\sigma$ [MeV] & pole position [MeV] & binding energy [MeV] & $x_{13}$ [$-$]\\\hline
   $610$& $1822.1-3.6i$ & $74.6-3.6i$ & $0.21$ \\
   $656$ & $1830.8-4.5i$ & $65.9-4.5i$ & $0.165$ \\
   $700$ & $1839.7-7.2i$ & $57.0-7.2i$ & $0.153$ \\
   $743$ & $1848.0-10.6i$ & $48.7-10.6i$ & $0.15$
  \end{tabular}
  \label{tab_pole_traj}
 \end{center}
\end{table}
The binding energy of the $\eta'N$ system is slightly
sensitive to the sigma meson mass, and is
roughly varied from $75$ to
$45$ MeV, and the imaginary part of the pole position from $-4$ to
$-10$ MeV with the change of $m_\sigma$ from 600 to 750 MeV.

\section{Summary and discussion\label{sec_summary}}
In this paper, we investigate the $\eta'N$ system using the three-flavor
linear
sigma model.
The $\eta'N$ interaction is evaluated using the tree-level amplitude
from the linear $\sigma$ model including the effect of the flavor
symmetry breaking.
The effect of the $\pi N$ channel is phenomenologically taken into
account
to reproduce the experimental data given in Ref.~\cite{Rader:1973mx}.
We also find an $\eta'N$ bound state
if the flavor symmetry breaking is taken into account;
the attractive interaction of $\eta'N$ becomes stronger
because the coupling of the sigma meson and nucleon becomes
stronger with the inclusion of the mixing effect.
The parameter $g$ which controls the strength of the coupling between the
sigma meson and nucleon is larger than that without the mixing effect.
In this study, the effect of the transition to the $\eta N$ and $\pi N$
channel does not seem to be so large because the imaginary part of the pole
is relatively small compared with its real part.
{The attractive $\eta'N$ interaction and the possible bound state
appearing from the sigma exchange contribution in association with the
U$_A$(1) anomaly would be suggestive for future theoretical
and experimental studies about the properties of the $\eta'$ meson.
In such studies,
a phenomenological approach based on
nonlinear sigma models
{respecting the experimental data with the inclusion of the $\pi N$
and $\eta N$ channels}
should be followed.

As we have commented in the text, the exchanges of the
scalar mesons cause the transitions of the $\eta'N$ into the other
channels.
There may be room for consideration of the inclusion of the
$K\Lambda$ and $K\Sigma$ channels; the transitions into these channels
can appear from the exchange of the strange scalar meson, e.g., the $\kappa$
meson in this model.
Then, the investigation of the possible effect from these channels
remains as a future task while we have concentrated on the non-strange
hadrons in this study.

The existence of such a bound state would have some impact on the
spectrum of the $\eta'$-nucleus system studied in
Refs.~\cite{Tsushima1,Tsushima2,Tsushima3,Nagahiro:2004qz,Nagahiro:2006dr};
the existence of the bound state generally gives the energy
dependence to the $\eta'$-optical potential in the nuclear medium.
Then, the expected spectrum of the $\eta'$-nucleus system studied in
Refs.~\cite{Tsushima1,Tsushima2,Tsushima3,Nagahiro:2004qz,Nagahiro:2006dr}
would be modified if one takes account of the possible $\eta'N$ bound state.

\section*{Acknowledgements}
 S.~S. was a JSPS fellow and appreciates the support of a JSPS
 Grant-in-Aid No.~25-1879.
 The work of D.~J. was partly supported by Grant-in-Aid for Scientific
 Research from JSPS 25400254.

\appendix
\section{Meson Mass}
\label{app_meson_mass}
The explicit form of the meson mass is given in this appendix.
The mass of the scalar and pseudoscalar mesons in the basis of the
Gell-Mann matrix are summarized in Table~\ref{tab_meson_mass}.
Here, we parameterize the squared meson mass $m^2$ as
\begin{align}
 m^2=c_\mu\mu^2+c_\lambda\lambda+c_{\lambda'}\lambda'+c_BB
\end{align}
and the
coefficients $c_\mu$, $c_\lambda$, $c_{\lambda'}$, and $c_B$ for each
meson are presented in the table.
\begin{table}
 \begin{center}
  \caption{Table of the meson mass.
  The squared mass of the meson $m^2$ is parameterized as
  $m^2=c_\mu\mu^2+c_\lambda\lambda+c_{\lambda'}\lambda'+c_BB$.}
  \begin{tabular}{c|cccc}
   meson&$c_\mu$&$c_\lambda$ &$c_{\lambda'}$ &$c_B$ \\\hline
   $m_{\sigma_0}^2$&$1$&$\so^2+\soo^2$ &$3\so^2+\soo^2$ &$-4\so$ \\
   $m_{\sigma_3}^2$&$1$&$(\so+\frac{\soo}{\sqrt{2}})^2$ & $\so^2+\soo^2$& $2(\so-\sqrt{2}\soo)$\\
   $m_{\sigma_8}^2$&$1$&$\so^2-\sqrt{2}\so\soo+\frac{3}{2}\soo^2$
       &$\so^2+3\soo^2$ &$2(\so+\sqrt{2}\soo)$ \\
   $m_{\sigma_0\sigma_8}^2$&$0$&$\frac{1}{2}\soo(4\so-\sqrt{2}\soo)$
       &$2\so\soo$ &$2\soo$ \\
   $m_{\eta_0}^2$&$1$&$\frac{1}{3}(\so^2+\soo^2)$& $\so^2+\soo^2$&$4\so$ \\
   $m_{\pi_3}^2$&$1$&$\frac{1}{3}(\so^2+\sqrt{2}\so\soo+\frac{\soo^2}{2})$
       &$\so^2+\soo^2$ &$-2(\so-\sqrt{2}\soo)$ \\ 
   $m_{\eta_8}^2$&$1$&$\frac{1}{3}(\so^2-\sqrt{2}\so\soo+\frac{3}{2}\soo^2)$ &$\so^2+\soo^2$ &$-2(\so+\sqrt{2}\soo)$ \\
   $m_{\eta_0\eta_8}^2$&$0$&$\frac{2}{3}\soo(\so-\frac{\soo}{2\sqrt{2}})$
       &$0$ &$-2\soo$
  \end{tabular}
 \end{center}
 \label{tab_meson_mass}
\end{table}
The transitions between the isovector and isoscalar mesons, $\sigma_3$
and $\sigma_{0,8}$ or $\pi_3$ and $\eta_{0,8}$, are suppressed by the
isospin symmetry.
The masses of the neutral mesons $\sigma,f_0$ and $\eta,\eta'$ are
obtained
as the eigenvalues of
the mass matrices defined in
Eqs.~(\ref{eq_sigma_mixing_mtrx}) and (\ref{eq_eta_mixing_mtrx}).
They are given as
$m^2_{\sigma/f_0}=\frac{1}{2}(m_{\sigma_0}^2+m_{\sigma_8}^2\mp\sqrt{(m_{\sigma_0}^2-m_{\sigma_8}^2)^2+4(m_{\sigma_0\sigma_8}^2)^2})$
and 
$m^2_{\eta'/\eta}=\frac{1}{2}(m_{\eta_0}^2+m_{\eta_8}^2\pm\sqrt{(m_{\eta_0}^2-m_{\eta_8}^2)^2+4(m_{\eta_0\eta_8}^2)^2})$.
Using the meson masses,
the mixing angles of the scalar and pseudoscalar sectors,
$\theta_s$ and $\theta_{ps}$, are given by $\tan
2\theta_s=2m^2_{\sigma_0\sigma_8}/(m^2_{\sigma_0}-m^2_{\sigma_8})$ and
$\tan 2\theta_{ps}=2m^2_{\eta_0\eta_8}/(m_{\eta_0}^2-m_{\eta_8}^2)$.

\section{Form of $V_{i\rightarrow j}$\label{app_vij}}
In this appendix, we present the explicit form of the interaction kernel
of the scattering equation.
As given in Eq.~(\ref{eq_proj_def}),
we apply the $s$-wave projection
and the average of the initial state nucleon.
The couplings $g_{\sigma_iN}$, $g_{\eta_iN}$, and $g_{\sigma_i\eta_j\eta_k}$
($\eta_i=\pi_3,\eta_0,\eta_8$ and $\sigma_i=\sigma_0,\sigma_3,\sigma_8$)
appearing in the following equations are given in
Appendix~\ref{app_coupling_mesons}.

Here, $p$, $k$, $p'$, and $k'$ are the four momenta of the initial-state
nucleon, the initial-state meson, the final-state nucleon, and the
final-state meson
in the center-of-mass frame which are given in
Eqs.~(\ref{eq_pmu})$-$(\ref{eq_kpmu}).
The matrix element of $\eta_iN\rightarrow\eta_jN$ in the tree level is
 written as follows:
\begin{align}
 \mathcal{M}_{\eta_iN\rightarrow\eta_jN}=&
 \bar{u}(p',s')\left[\frac{g_{\sigma_k\eta_i\eta_j}g_{\sigma_kN}}{q^2-m_{\sigma_k}^2+i\epsilon}+\frac{g_{\eta_iN}g_{\eta_jN}\Slash{k}}{(p+k)^2-m_N^2+i\epsilon}-\frac{g_{iN}g_{jN}\Slash{k}'}{(p-k')^2-m_N^2+i\epsilon}\right]u(p,s).\label{app_eq_mat_elem_ij}
\end{align}
First, $\frac{1}{2}\sum_{s,s'}\bar{u}(p',s')\Slash{k}u(p,s)$ is written as
 \begin{align}
  &\frac{1}{2}\sum_{s,s'}\bar{u}(p',s')\Slash{k}u(p,s)
  =\frac{\sqrt{(E_N+m_N)(E_N'+m_N)}}{2m_N}\notag\\
  &\hspace{1cm}\cdot\left[E_i+\frac{\pcm^2}{E_N+m_N}+(E_i+E_N+m_N)\frac{\pcm\pcm'\cos\theta}{(E_N+m_N)(E_N'+m_N)}\right], 
 \end{align}
 where the Dirac spinor $u(p,s)$ is normalized as
 $\bar{u}(p,s')u(p,s)=\delta_{ss'}$ and $\chi_s$ is the Pauli spinor
 normalized as $\chi_{s'}^\dagger\chi_s=\delta_{ss'}$.
 In the same way, $\frac{1}{2}\sum_{s,s'}\bar{u}(p',s')\Slash{k}'u(p,s)$ is written as
 \begin{align}
  &\frac{1}{2}\sum_{s,s'}\bar{u}(p',s')\Slash{k}'u(p,s)
  =\frac{\sqrt{(E_N+m_N)(E_N'+m_N)}}{2m_N}\notag\\
  &\hspace{1cm}\cdot\left[E_j+\frac{\pcm'^2}{E_N'+m_N}+(E_j+E_N'+m_N)\frac{\pcm\pcm'\cos\theta}{(E_N+m_N)(E_N'+m_N)}\right].
 \end{align}
Moreover, $\frac{1}{2}\sum_{s,s'}\bar{u}(p,s')u(p,s)$ is written as
 \begin{align}
  \frac{1}{2}\sum_{s,s'}\bar{u}(p',s')u(p,s)
  =&\frac{\sqrt{(E_N+m_N)(E_N'+m_N)}}{2m_N}\left(1-\frac{\pcm\pcm'\cos\theta}{(E_N+m_N)(E_N'+m_N)}\right).
 \end{align}
 Substituting the explicit form of $p$, $k$, $p'$, and $k'$ in
 Eqs.~(\ref{eq_pmu})$-$(\ref{eq_kpmu}), the matrix element is written as
 \begin{align}
 &\frac{1}{2}\sum_{s,s'}\mathcal{M}_{\eta_iN\rightarrow \eta_jN}=\frac{\sqrt{(E_N+m_N)(E_N'+m_N)}}{2m_N}\left[\sum_{\sigma_k}\frac{g_{\sigma_kN}g_{\sigma_k\eta_i\eta_j}}{2\pcm\pcm'\cos\theta-m_{\sigma_k}^2+2m_N^2-2E_NE_N'}\right.\notag\\
  &\hspace{3mm}+\frac{g_{\eta_iN}g_{\tau_jN}}{W^2-m_N^2+i\epsilon}\left(E_i+\frac{\pcm^2}{E_N+m_N}+(E_i+E_N+m_N)\frac{\pcm\pcm'\cos\theta}{(E_N+m_N)(E_N'+m_N)}\right)\notag\\
  &\hspace{3mm}+\frac{g_{\eta_iN}g_{\eta_jN}}{2(E_NE_j+\pcm\pcm'\cos\theta)-m_j^2}\left(E_j+\frac{\pcm'^2}{E_N'+m_N}\right.\notag\\
  &\left.\left.\hspace{3mm}+(E_j+E_N'+m_N)\frac{\pcm\pcm'\cos\theta}{(E_N+m_N)(E_N'+m_N)}\right)\right].
 \end{align}
With the $s$-wave projection $\frac{1}{2}\int_{-1}^1d\cos\theta$,
the matrix element is written as
 \begin{align}
  &\frac{1}{2}\int^1_{-1}d(\cos\theta)\frac{1}{2}\sum_{s,s'}\mathcal{M}_{\eta_iN\rightarrow
  \eta_jN}=\frac{\sqrt{(E_N+m_N)(E_N'+m_N)}}{2m_N}\left[\sum_{\sigma_k}\left\{-\frac{1}{2(E_N'+m_N)(E_N+m_N)}\right.\right.\notag\\
  &\left.\hspace{3mm}+\frac{1}{4\pcm\pcm'}\left(1-\frac{m_{\sigma_k}^2-2m_N^2+2E_NE_N'}{2(E_N+m_N)(E_N'+m_N)}\right)\ln\left(\frac{2\pcm\pcm'-m_{\sigma_k}^2+2m_N^2-2E_NE_N'}{-2\pcm\pcm'-m_{\sigma_k}^2+2m_N^2-2E_NE_N'}\right)\right\}\notag\\
  &\hspace{3mm}+\frac{g_{\eta_iN}g_{\eta_jN}}{W^2-m_N^2}\left(E_i+\frac{\pcm^2}{E_N+m_N}\right)+g_{\eta_iN}g_{\eta_jN}\left\{\frac{W+m_N}{2(E_N+m_N)(E_N'+m_N)}\right.\notag\\
  &\hspace{3mm}+\frac{1}{4\pcm\pcm'}\left(E_j+\frac{\pcm'^2}{E_N'+m_N}-\frac{(W+m_N)(2E_NE_j-m_j^2)}{2(E_N+m_N)(E_N'+m_N)}\right)\notag\\
  &\left.\left.\hspace{6mm}\cdot\ln\left(\frac{2\pcm\pcm'+2E_NE_j-m_j^2}{-2\pcm\pcm'+2E_NE_j-m_j^2}\right)\right\}\right]. 
 \end{align}
Then, the explicit form of $V_{i\rightarrow j}$ $(i,j=\eta'N$, $\eta N$,
or $\pi N$) is:
\begin{align}
 &V_{\eta'N\rightarrow\eta'N}=-\sum_{\sigma_k}\frac{g_{\eta'N}g_{\eta'\eta'\sigma_k}}{m_{\sigma_k}^2}+\frac{4m_Ng_{\eta'N}^2}{4m_N^2-m_{\eta'}^2},\\
 &V_{\eta_iN\rightarrow\eta_i
 N}=\frac{E_N+m_N}{2m_N}\left[\sum_{\sigma_k}g_{\sigma_k\eta_i\eta_i}g_{\sigma_kN}\left\{-\frac{1}{2(E_N+m_N)^2}+\frac{1}{4\pcm'^2}\left(1-\frac{m_{\sigma_k}^2+2\pcm^2}{2(E_N+m_N)^2}\right)\right.\right.\notag\\
 &\hspace{6mm}\cdot\ln\left(\frac{m_{\sigma_k}^2}{m_{\sigma_k}^2+4\pcm^2}\right)
 +\frac{g_{\eta_i
 N}^2\left(E_i+\frac{\pcm^2}{E_N+m_N}\right)}{W^2-m_N^2}+g_{\eta_i
 N}^2\left\{\frac{W+m_N}{2(E_N+m_N)^2}+\frac{1}{4\pcm^2}\right.\notag\\
 &\hspace{6mm}\cdot\left(E_i+\frac{\pcm^2}{E_N+m_N}-\frac{(W+m_N)(2E_NE_i-m_i^2)}{2(E_N+m_N)^2}\right)\notag\\
 &\left.\left.\hspace{6mm}\cdot\ln\left(\frac{2\pcm^2+2E_NE_i-m_i^2}{-2\pcm^2+2E_NE_i-m_i^2}\right)\right\}\right]\
 (\eta_i=\pi^0,\eta),\\
 &V_{\pi^+n\rightarrow\pi^+n}=
 \frac{E_N+m_N}{2m_N}\left[\sum_{\sigma_k}g_{\sigma_k\pi^+\pi^-}g_{\sigma_kN}\left\{-\frac{1}{2(E_N+m_N)^2}\right.\right.\notag\\
 &\left.\left.\hspace{3mm}+\frac{1}{4\pcm'^2}\left(1-\frac{m_{\sigma_k}^2+2\pcm^2}{2(E_N+m_N)^2}\ln\left(\frac{m_{\sigma_k}^2}{m_{\sigma_k}^2+4\pcm^2}\right)\right)
 \right\}+\frac{g_{\pi^+n}^2\left(E_i+\frac{\pcm^2}{E_N+m_N}\right)}{W^2-m_N^2}\right],\\
 &V_{\pi^+p\rightarrow\pi^+p}=
 \frac{E_N+m_N}{2m_N}\left[\sum_{\sigma_k}g_{\sigma_k\pi^+\pi^-}g_{\sigma_kN}\left\{-\frac{1}{2(E_N+m_N)^2}\right.\right.\notag\\
 &\left.\hspace{3mm}+\frac{1}{4\pcm'^2}\left(1-\frac{m_{\sigma_k}^2+2\pcm^2}{2(E_N+m_N)^2}\ln\left(\frac{m_{\sigma_k}^2}{m_{\sigma_k}^2+4\pcm^2}\right)\right)\right\}
 +g_{\pi^+p}^2\left\{\frac{W+m_N}{2(E_N+m_N)}\right.\notag\\
 &\hspace{3mm}+\frac{1}{4\pcm^2}\left(E_i+\frac{\pcm^2}{E_N+m_N}-\frac{(W+m_N)(2E_NE_i-m_i^2)}{2(E_N+m_N)^2}\right)\notag\\
 &\left.\left.\hspace{6mm}\cdot\ln\left(\frac{2\pcm^2+2E_NE_i-m_i^2}{-2\pcm^2+2E_NE_i-m_i^2}\right)\right\}\right], \\ 
 &V_{\eta' N\rightarrow\eta_j
 N}=\sqrt{\frac{E_N'+m_N}{2m_N}}\left[\sum_{\sigma_k}g_{\eta'N}g_{\eta'\eta_j\sigma_k}\left\{-\frac{1}{4m_N(E_N'+m_N)}+\frac{1-\frac{m_{\sigma_k}^2-2m_N^2+2m_NE_N'}{4m_N(E_N'+m_N)}}{2m_N^2-m_{\sigma_k}^2-2m_NE_N'}\right\}\right.\notag\\
 &\left.\hspace{3mm}+\frac{g_{\eta'N}g_{\eta_jN}}{2m_N+m_{\eta'}}+g_{\eta'N}g_{\eta_jN}\frac{\left(E_j+\frac{\pcm'^2}{E_N'+m_N}\right)}{2m_NE_j-m_j^2}\right]
 (\eta_j=\pi^0,\eta),\\
 &V_{\pi^+n\rightarrow\pi^0p}=\frac{E_N+m_N}{2m_N}\left[\sum_{\sigma_k}g_{\sigma_k\pi^+\pi^0}g_{\sigma_kN}\left\{-\frac{1}{2(E_N+m_N)^2}\right.\right.\notag\\
 &\left.\left.\hspace{3mm}+\frac{1}{4\pcm'^2}\left(1-\frac{m_{\sigma_k}^2+2\pcm^2}{2(E_N+m_N)^2}\right)\ln\left(\frac{m_{\sigma_k}^2}{m_{\sigma_k}^2+4\pcm^2}\right)
 \right\}+\frac{g_{\pi^+N}g_{\pi^0p}\left(E_i+\frac{\pcm^2}{E_N+m_N}\right)}{W^2-m_N^2}\right.\notag\\
 &\hspace{3mm}+g_{\pi^+N}g_{\pi^0n}\left\{\frac{W+m_N}{2(E_N+m_N)^2}+\frac{1}{4\pcm^2}\left(E_i+\frac{\pcm^2}{E_N+m_N}-\frac{(W+m_N)(2E_NE_i-m_i^2)}{2(E_N+m_N)^2}\right)\right.\notag\\
 &\left.\left.\hspace{6mm}\cdot\ln\left(\frac{2\pcm^2+2E_NE_i-m_i^2}{-2\pcm^2+2E_NE_i-m_i^2}\right)\right\}\right],\\
 &V_{\eta N\rightarrow\pi
 N}=\frac{\sqrt{(E_N+m_N)(E_N'+m_N)}}{2m_N}\left[\sum_{\sigma_k}\left\{-\frac{1}{2(E_N+m_N)(E_N'+m_N)}\notag\right.\right.\\
 &\left.\hspace{3mm}+\frac{1}{4\pcm\pcm'}\left(1-\frac{m_{\sigma_k}^2-2m_N^2+2E_NE_N'}{2(E_N+m_N)(E_N'+m_N)}\right)\cdot
 \ln\left(\frac{2\pcm\pcm'-m_{\sigma_k}^2+2m_N^2-2E_NE_N'}{-2\pcm\pcm'-m_{\sigma_k}^2+2m_N^2-2E_NE_N'}\right)\right\}\notag\\
 &\left.\hspace{3mm}+\frac{g_{\eta N}g_{\pi
 N}}{W^2-m_N^2}\left(E_i+\frac{\pcm^2}{E_N+m_N}\right)+g_{\eta N}g_{\pi
 N}\left\{\frac{W+m_N}{2(E_N+m_N)(E_N'+m_N)}+\frac{1}{4\pcm\pcm'}\right.\right.\notag\\
 &\left.\left.\hspace{6mm}\cdot\left(E_j+\frac{\pcm'^2}{E_N'+m_N}-\frac{(W+m_N)(2E_NE_j-m_j^2)}{2(E_N+m_N)(E_N'+m_N)}\right)\ln\left(\frac{2\pcm\pcm'+2E_NE_N'-m_j^2}{-2\pcm\pcm'+2E_NE_j-m_j^2}\right)\right\}\right].
\end{align}
The transition term $V_{\eorep p\rightarrow\pi^+n}$ satisfies
$V_{\eorep p\rightarrow\pi^+n}=V_{\eorep
 n\rightarrow\pi^-p}=\sqrt{2}V_{\eorep
 p\rightarrow\pi^0p}=-\sqrt{2}V_{\eorep n\rightarrow\pi^0n}$.
 The transition term from $\eorep N$ to $\pi N$ $(I=1/2)$ is given as
 \begin{align}
  V_{\eorep N\rightarrow\pi
  N(I=1/2,I_z=1/2)}=&\frac{1}{\sqrt{3}}\left(V_{\eorep
  p\rightarrow\pi^0p}+\sqrt{2}V_{\eorep p\rightarrow\pi^+n}\right)\\
  V_{\eorep N\rightarrow\pi
  N(I=1/2,I_z=-1/2)}=&-\frac{1}{\sqrt{3}}\left(V_{\eorep
  n\rightarrow\pi^0n}-\sqrt{2}V_{\eorep n\rightarrow\pi^-p}\right)
 \end{align}
Then, $V_{\eorep N\rightarrow\pi N(I=1/2)}$ is given as
$V_{\eorep N\rightarrow\pi N(I=1/2,I_z=\pm 1/2)}=\sqrt{3}V_{\eta'N\rightarrow\pi^0N}$.
For the $\pi N$ $(I=1/2)$ elastic channel, the matrix element is written
as
\begin{align}
 V_{\pi N\rightarrow\pi N(I=1/2,I_z=1/2)}=&\frac{1}{3}\left(V_{\pi^0p\rightarrow\pi^0p}+\sqrt{2}V_{\pi^0p\rightarrow\pi^+n}+\sqrt{2}V_{\pi^+n\rightarrow\pi^0p}+2V_{\pi^+n\rightarrow\pi^+n}\right),\\
 V_{\pi N\rightarrow\pi N(I=1/2,I_z=-1/2)}=&\frac{1}{3}\left(V_{\pi^0n\rightarrow\pi^0n}-\sqrt{2}V_{\pi^0n\rightarrow\pi^-p}-\sqrt{2}V_{\pi^-p\rightarrow\pi^0n}+2V_{\pi^-p\rightarrow\pi^-p}\right).
\end{align}
Here, $V_{\pi^0p\rightarrow\pi^+n}=-V_{\pi^0n\rightarrow\pi^-p}$,
$V_{\pi^0n\rightarrow\pi^0n}=V_{\pi^0p\rightarrow\pi^0p}$, and
$V_{\pi^+n\rightarrow\pi^+n}=V_{\pi^-p\rightarrow\pi^-p}$.
Then, $V_{\pi N\rightarrow\pi N(I=1/2,I_z=1/2)}=V_{\pi N\rightarrow\pi
 N(I=1/2,I_z=-1/2)}$ is satisfied.

\section{Couplings of hadrons}
\label{app_coupling_mesons}
In this appendix, we present the couplings of the hadrons appearing in
the text.
The couplings of the meson ($\sigma_i$ or $\eta_i$) and nucleon are
summarized in Table~\ref{tab_coup_mb}.
$g_{\sigma_iN}$ and $g_{\eta_iN}$ denote the coefficients of the terms
$\bar{N}\sigma_iN$ and $\bar{N}\eta_i\gamma_5N$ in the Lagrangian
Eq.~(\ref{eq_lag}), respectively.
\begin{table}
 \centering
 \caption{Meson$-$baryon couplings.
 $\tau_3$ is the third component of the Pauli matrix.}
 \begin{tabular}{cccccccc}
  $g_{\sigma_0N}$&$g_{\sigma_3N}$&$g_{\sigma_8N}$&$g_{a_0^\pm
	      N}$&$g_{\eta_0N}$&$g_{\pi_3N}$&$g_{\eta_8N}$&$g_{\pi^\pm
			      N}$\\\hline
  $-g/\sqrt{3}$&$-g\tau_3/\sqrt{2}$ &$-g/\sqrt{6}$ &$-g$&$g/\sqrt{3}$ &$g\tau_3/\sqrt{2}$ &$g/\sqrt{6}$&$g$ \\
 \end{tabular}
 \label{tab_coup_mb}
\end{table}
The couplings of the mesons $g_{\sigma_i\eta_j\eta_k}$, which are
defined as the coefficients of the term $\sigma_i\eta_j\eta_k$ in the
Lagrangian, are given in Table~\ref{tab_coup_mesons}.
The parameters $a$, $b$, and $c$ in the table characterize the coupling
of the meson as
\begin{align}
 g_{\sigma_i\eta_j\eta_k}=-(a\lambda+b\lambda'+cB).
\end{align}
\begin{table}
 \centering
 \caption{Meson couplings.
 $a$, $b$, and $c$ are defined by
 $g_{\sigma_i\eta_j\eta_k}=-(a\lambda+b\lambda'+cB)$, where $\lambda$,
 $\lambda'$, and $B$ are the parameters appearing in the Lagrangian.}
 \begin{tabular}{c|ccc}
  &$a$ &$b$ &$c$ \\\hline
  $g_{\sigma_0\eta_0\eta_0}$&$\frac{2}{3}\so$ &$2\so$ &$4$ \\
  $g_{\sigma_8\eta_0\eta_0}$&$\frac{2}{3}\soo$ &$2\soo$ &$0$ \\
  $g_{\sigma_0\pi_3\pi_3}$&$\frac{2}{3}(\so+\frac{\soo}{\sqrt{2}})$
      &$2\so$ &$-2$ \\
  $g_{\sigma_3\pi_3\pi_3}$&$0$	&$0$	&$0$	\\
  $g_{\sigma_8\pi_3\pi_3}$&$\frac{\sqrt{2}}{3}(\so+\frac{\soo}{\sqrt{2}})$
      &$2\soo$ &$2\sqrt{2}$	\\
  $g_{a_0^\pm\pi_3\pi^\mp}$&$0$	&$0$	&$0$	\\
  $g_{\sigma_0\eta_8\eta_8}$&$\frac{2}{3}(\so-\frac{\soo}{\sqrt{2}})$
      &$2\so$	&$-2$	\\
  $g_{\sigma_8\eta_8\eta_8}$&$-\frac{\sqrt{2}}{3}(\so-\frac{3}{\sqrt{2}}\soo)$&$2\soo$&$-2\sqrt{2}$	\\
  $g_{\sigma_0\eta_0\eta_8}$&$\frac{2}{3}\soo$	&$0$
	  &$0$	\\
  $g_{\sigma_8\eta_0\eta_8}$&$\frac{2}{3}(\so-\frac{\soo}{\sqrt{2}})$	&$0$
	  &$-2$	\\
  $g_{\sigma_3\eta_0\pi_3}$&$\frac{2}{3}(\so+\frac{\soo}{\sqrt{2}})$
      &$0$&$-2$	\\
  $g_{\sigma_3\pi_3\eta_8}$&$\frac{\sqrt{2}}{3}(\so+\frac{\soo}{\sqrt{2}})$
      &$0$	&$2\sqrt{2}$	\\
 \end{tabular}
 \label{tab_coup_mesons}
\end{table}
 Due to the isospin symmetry,
 $g_{\sigma_{0,3,8}\pi^\pm\pi^\mp}=g_{\sigma_{0,3,8}\pi_3\pi_3}$,
 $g_{a_0^\pm\eta_0\pi^\mp}=g_{\sigma_3\eta_0\pi_3}$, and
 $g_{a_0^\pm\pi^\mp\eta_8}=g_{\sigma_3\pi_3\eta_8}$,
 and the couplings $g_{\sigma_3\eta_{0,8}\eta_{0,8}}$ and
 $g_{\sigma_{0,8}\pi_3\eta_{0,8}}$ vanish.
 Here, $\eta_i$ and $\sigma_i$ are the eigenstates of the Gell-Mann
 matrix appearing in the Lagrangian,
 and these states are mixed by the flavor SU(3) symmetry breaking.
 The eigenstates of the mass $\sigma$, $f_0$, $\eta'$, and $\eta$ and
 the Gell-Mann matrix $\sigma_i$ and $\eta_i$ ($i=0,8$) are related by
 the matrices in Eqs.~(\ref{eq_sigma_mixing_mtrx}) and
 (\ref{eq_eta_mixing_mtrx}), respectively.
 The coupling of $\sigma$ ($\eta'$) and $f_0$ ($\eta$) to the hadronic
 state $h$ in the mass eigenstate is obtained using the mixing angles
 $\theta_s$ and $\theta_{ps}$ for the scalar and pseudoscalar mesons:
 \begin{align}
  g_{\sigma(\eta')
  h}=&\cos\theta_{s(ps)}g_{\sigma_0(\eta_0)h}+\sin\theta_{s(ps)}g_{\sigma_8(\eta_8)h},\\
  g_{f_0(\eta) h}=&-\sin\theta_{s(ps)}g_{\sigma_0(\eta_0)h}+\cos\theta_{s(ps)}g_{\sigma_8(\eta_8)h}.
 \end{align}


\end{document}